%% file: main.tex
\documentclass[10pt,conference]{IEEEtran}
\usepackage{cite}
\usepackage{amsmath,amssymb,amsfonts}
\usepackage{algorithmic}
\usepackage{graphicx}
\usepackage{textcomp}
\usepackage{xcolor}
\usepackage{comment}
\usepackage{tikz}
\usepackage{hyperref}

\def\BibTeX{{\rm B\kern-.05em{\sc i\kern-.025em b}\kern-.08em
    T\kern-.1667em\lower.7ex\hbox{E}\kern-.125emX}}

\definecolor{codeColor}{RGB}{139,26,26}

\newcommand{\ali}[1]{\textcolor{blue}{{\it Ali: #1}}}
\newcommand{\code}[1]{\textcolor{codeColor}{\textsf{#1}}}

\newcommand{\figref}[1]{Figure~\ref{#1}}
\newcommand{\secref}[1]{Section~\ref{#1}}

\newcommand*\circled[1]{\tikz[baseline=(char.base)]{
\node[shape=circle,font=\bfseries,thick,draw=black,fill=black,text=white,inner sep=0.8pt] (char) {#1};}}

\hypersetup{
    colorlinks=true,
    linkcolor=blue,
    filecolor=blue,      
    urlcolor=blue,
} 

\begin{document}

%%
%% The "title" command has an optional parameter,
%% allowing the author to define a "short title" to be used in page headers.
\title{DepRes: A Tool for Resolving Fully Qualified Names and Their Dependencies}
%\title{Supporting Program Analysis Tools by ResolvingProgram Dependencies in Code Repositories}

\author{
\IEEEauthorblockN{1\textsuperscript{st} Ali Shokri}
\IEEEauthorblockA{\textit{Global Cybersecurity Institute} \\
\textit{Rochester Institute of Technology}\\
Rochester, USA \\
as8308@rit.edu}
\and
\IEEEauthorblockN{2\textsuperscript{nd} Mehdi Mirakhorli}
\IEEEauthorblockA{\textit{Global Cybersecurity Institute} \\
\textit{Rochester Institute of Technology}\\
Rochester, USA \\
mxmvse@rit.edu}
}

\maketitle  

%%
%% The abstract is a short summary of the work to be presented in the
%% article.
\begin{abstract}
Reusing code snippets shared by other programmers on Q\&A forums (e.g., StackOverflow) is a common practice followed by software developers. However, lack of sufficient information about the fully qualified name (FQN) of identifiers in borrowed code snippets, results in serious compile errors. Programmers either have to manually search for the correct FQN of identifiers which is a tedious and error-prone process, or use tools developed to automatically identify correct FQNs. Despite the efforts made by researchers to automatically identify FQNs in code snippets, the current approaches suffer from low accuracy when it comes to practice. Moreover, while these tools focus on resolving the FQN for an identifier in a code snippet, they leave the challenge of finding the correct third-party library (i.e., dependency) implementing that FQN unresolved. Using an incorrect dependency or incorrect version of a dependency might lead to a semantic error which is not detectable by compilers. Therefore, it can result in serious damage in the run-time.

In this paper, we introduce DepRes, a tool that leverages a sketch-based approach to resolve FQNs in java-based code snippets and recommend the correct dependency for each FQN. The source code, documentation, as well as a demo video of DepRes tool is available from \url{https://github.com/SoftwareDesignLab/DepRes-Tool}.

 % Software developers increasingly rely on cloning projects from open-source code repositories (e.g., GitHub) as a basis to build their software upon. Moreover, reusing code snippets shared by other programmers on Q\&A forums (e.g., StackOverflow) is a common practice followed by software developers. However, lack of proper definition of fully-qualified names of types in borrowed code snippets, as well as insufficient third-party dependency declaration in cloned projects results in serious compile errors. Programmers either have to manually search the correct FQN of types which is a tedious and error-prone process, or use some tools developed to automatically identify correct FQNs. Despite the efforts made by researchers to automatically identifying FQNs in code snippets, the current approaches suffer from low accuracy when it comes to practice. Moreover, these tools focus on resolving FQNs and leave the challenge of finding correct dependencies in programs unresolved. This could result in serious semantic errors in the developed software.   
  
  %In this paper, we introduce DepRes, a tool that facilitates resolving projects' dependencies as well as identifying correct FQNs. Leveraging itemset mining and sketch-based programming, this tool is able to detect the correct FQNs for identifiers in a given code snippet, and recommend necessary dependencies needed to be added to the project. The source code, documentation, as well as a demo video of DepRes tool is available from \url{https://github.com/SoftwareDesignLab/DepRes-Plugin}.  
 \end{abstract}


\input{Introduction}

\input{DepRes}
\input{DepResInPractice}

\input{Limitations}
\input{Conclusion}

\bibliographystyle{IEEEtran}
\bibliography{IEEEabrv,Bibliography}

\end{document}

%% file: Introduction.tex
\section{Introduction}
\label{section:Introduction}

\begin{figure}[ht]
\centering
\includegraphics[width=.5\textwidth]{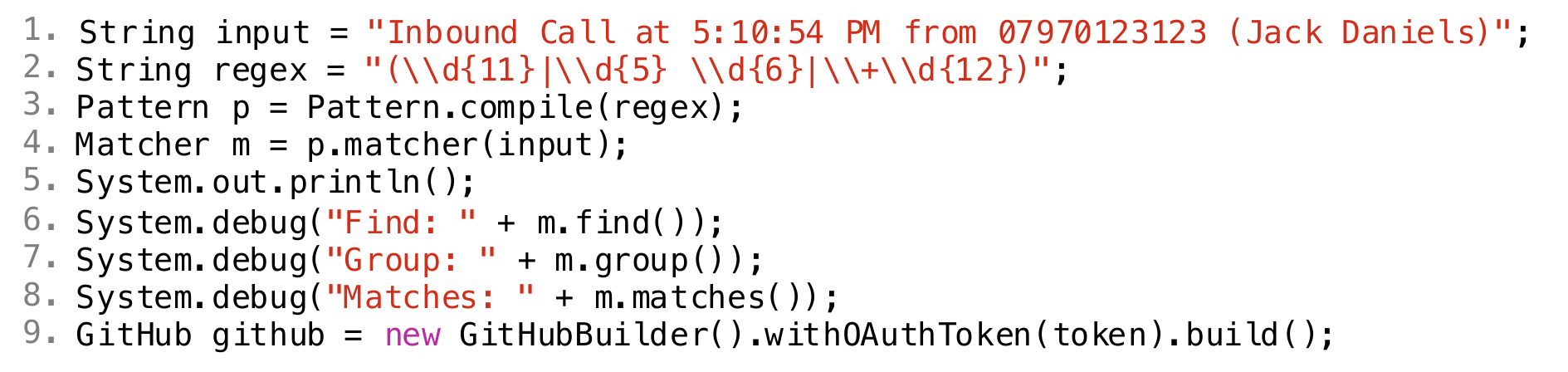}
\caption{An example of a code snippet without FQNs} 
\label{fig:sample_1}
\end{figure}

Borrowing a code snippet shared on an open-source code repository (e.g., GitHub\footnote{https://github.com}) or a Q\&A forum (e.g., Stackoverflow\footnote{https://stackoverflow.com}) is a common practice followed by software developers \cite{gharehyazie2017some, abdalkareem2017code}. The application of the reused code varies from building a new software \cite{kim2005empirical}, to investigating properties of the code snippet such as their API usage models \cite{shokri2021arcode}, or even training a model for correct code retrieval \cite{lv2015codehow}.
%Sharing the source code of a software on an open-source code repository (e.g., GitHub\footnote{https://github.com}) and letting other software developers to clone it is a common practice followed by programmers \cite{gharehyazie2017some}. The cloned software can be used in a variety of applications, from building a new software \cite{X} to investigating properties of the cloned software such as API usage models by performing different software analysis techniques \cite{shokri2021arcode}. In addition to borrowing the whole project from code repositories, programmers also share their questions and concerns over Q\&A forums (e.g., Stackoverflow\footnote{https://stackoverflow.com}), and reuse the code snippets included in answers provided by experts \cite{abdalkareem2017code}. 
While these code reuse approaches aim to save time, efforts, and avoid re-inventing the wheel, lack of providing \textit{fully qualified names (FQNs)} of identifiers in the shared code snippets, as well as insufficient \textit{dependency} declaration in the shared projects are serious barriers in achieving an accurate and proper outcome \cite{phan2018statistical}. Based on the role of an identifier in a code snippet, we consider two types of FQNs.
\begin{itemize}
    \item FQN for a type (e.g., a class in an object oriented language): the full packaging name followed by the simple name of the type;
     \item FQN for a method: the full signature of the method;
    % \item FQN for a field: the full packaging name followed by the simple name of its type followed by the name of the field.
\end{itemize}
For instance, in the code snippet shown in \figref{fig:sample_1}, a possible FQN of both identifiers \code{Pattern} and \code{p} in line 3 could be \code{java.util.regex.Pattern}. In such a case, the FQN of \code{compile} in the same line would be \code{java.util.regex.Pattern.compile(java.util.String)void}. It means that \code{compile} is a method declared in class \code{java.util.regex.Pattern}, with a single argument of type \code{java.lang.String}, which returns nothing (i.e., \code{void}). A programmer needs to know the FQN of each identifier in the code snippet to be able to add proper \textit{import}s to the program. Otherwise, a compile error will be raised. A \textbf{dependency}, on the other hand, refers to a third party library that actually implements an FQN. Please note that each FQN can be introduced in different libraries or even in different \textit{version} of a same library and hence, have different implementations. For instance, \code{java.util.regex.Pattern} is available in different versions of java JDK libraries. Adding an incorrect library or even an incorrect version of the expected library to the project can result in semantic errors in a program. These types of error are not necessarily detectable by compilers and might lead to run-time crashes. However, selecting the correct dependency and adding it to the project is not a trivial task. 

\begin{figure*}
\center
\includegraphics[width=\textwidth]{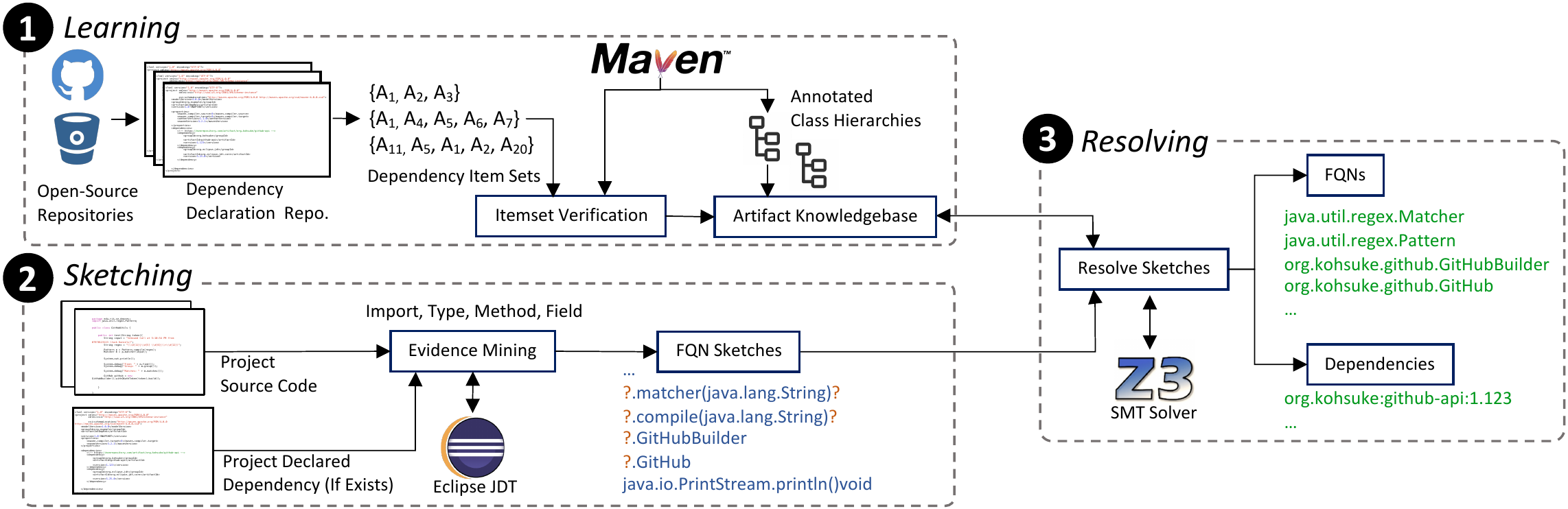}
\caption{An overview of DepRes three steps and their interactions} 
\label{fig:depres_overview}
\end{figure*}

\begin{comment}
Program analysis has found many applications from identifying errors (security, memory leak, etc.) in programs to mine a repository of code and create API usage patterns \cite{shokri2021arcode}.

Program Analysis from source can not catch the pace of new programming language features introduced. For instance, handling Lambda expressions by WALA, SOOTE, ...

Therefore, people need to use compiled version of the program (e.g. bytecode). \ali{Or even since different versions of libraries have different implementations (bugs fixed, new issues happen, etc.), if they use source code, they need to provide the complete list of dependencies with the \textbf{correct version} to have an accurate program analysis. }

To generate bytecodes for a repository of programs, programmers need to manually compile, inspect, and add dependencies. Even in this approach, finding a proper dependency in a program that is written by someone else (and you have no idea about the correct dependency) is a tedious as well as error-prone task. 
\end{comment}

%\ali{Add some works from ASE and rewrite this part}

There are several approaches introduced by researchers that aim to find correct FQN for an identifier in a code snippet. Baker \cite{subramanian2014live} creates an incomplete AST of a program and through an iterative process called \textit{deductive linking}, gradually finds FQNs. It needs an oracle to guide the process. Based on a corpus of programs in a code repository, StatType \cite{phan2018statistical} learns co-occurrence of FQNs in a scope. It then analyzes the code snippet under consideration and recommends FQNs based on the context of the identifier. In another line of work, researchers developed techniques for finding correct dependencies for FQNs. In \cite{ossher2010automated} authors aim to automatically identify the correct dependency through searching in global search engines and their database. Though they do not consider versioning of libraries, neither do they differentiate between different libraries that implement the same FQN. COSTER \cite{saifullah2019learning}, the state of the art FQN finder, learns local and global context of identifiers in a program. It then, searches through the FQNs in its database and finds the FQNs with the similar context to identifier's. Although this approach suggests the FQN, it can not find a correct version of the library. Also, developers do not always follow a specific naming convention for naming variables in their programs. Therefore, relying on token-based similarity metrics might not result in accurate recommendations.

Despite all the efforts made by researchers to address the mentioned challenges, these works suffer from a twofold issue. First, a relatively low accuracy in finding FQNs. Second, they do not consider the correct dependency or it's version.

To address the mentioned challenges, in this paper, we introduce DepRes, a tool that learns from two sources, projects in a code repository (e.g., GitHub), and a partial dependency provider (e.g., Maven). It then generates sketches of desired FQNs, and resolves the generated sketches. Software developers can use this tool to identify the fully qualified name of identifiers in a code snippet and find their corresponding dependencies based on the context of the project.

\begin{comment}
In this paper, we introduce DepRes to support programming analysis tools with automatically identifying missed or incorrect dependencies in programs that cause compilation failures. DepRes is able to automatically make changes to the list of programs' dependencies, compile them, and generate programs' jar files. The output of DepRes is a repository of jar files of compiled programs that include their dependencies. Based on our preliminary results, ...

\end{comment}

%% file: DepRes.tex
\begin{figure*}
\center
\includegraphics[width=\textwidth]{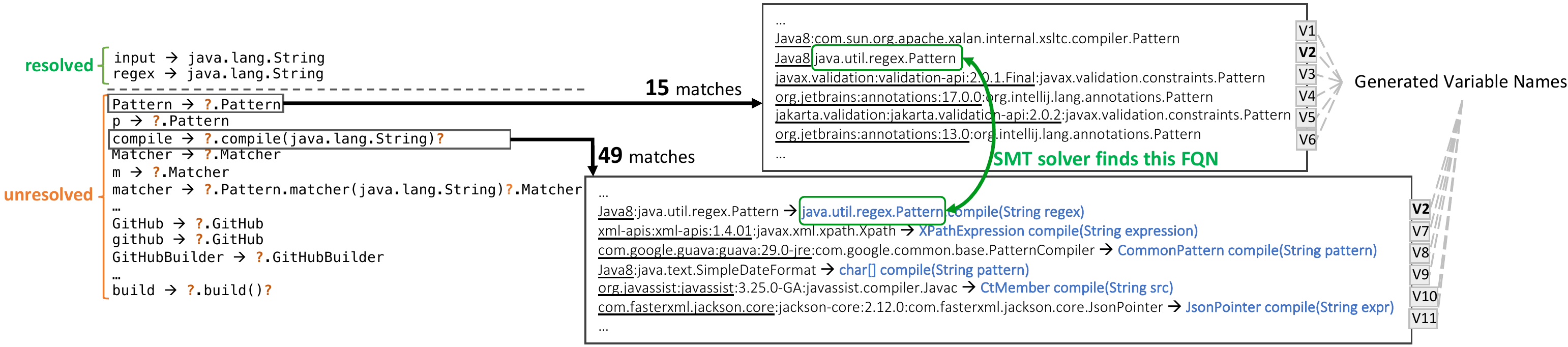}
\caption{A walkthrough example of DepRes performance} 
\label{fig:depres_in_practice}
\end{figure*}

\section{DepRes}
\label{section:DepRes}

\textbf{DepRes} is a tool that automatically resolves FQNs and project dependencies in a given code snippet. To that end, in a three step process, DepRes \textbf{(1)} \textbf{learns} from two sources, an open-source code repository, and a dependency repository to create a \textit{knowledgebase} of FQNs and their related dependencies, \textbf{(2)} analyzes the given code snippet and collects \textit{evidences} around identifiers and generates their corresponding \textbf{sketches}, and \textbf{(3)} \textbf{resolves} the created sketches. \figref{fig:depres_overview} demonstrates these steps. Following, we briefly explain each step. A walkthrough example is provided in \secref{section:DepResInPractice}.
\begin{itemize}[]

\item[\circled{1}]  \textbf{Learning}: 
In this step, DepRes automatically finds open-source projects in the code repository (e.g., GitHub), collects their dependency information from the corresponding file (e.g., POM.xml, build.gradle, .classpath), and creates an itemset of dependencies for each project. Then, for each dependency, DepRes downloads its jar file from Maven repository, creates the class hierarchy for that jar file, and annotates each class, method, and field in the class hierarchy with the name and version of the dependency. Maven also provides partial information about each dependency. This information includes some of the compile-time as well as test-time dependencies for each library. To make sure that the final knowledgebase contains only correct dependency information, DepRes uses the partial dependency information provided by Maven as a ground-truth and rules out dependency information gathered from the code repository that violate the information collected from Maven.  

\item[\circled{2}]  \textbf{Sketching}: 
Next, DepRes collects evidences for all the identifiers in the given code snippet. If the given code is not in the format of a project, DepRes creates a temporary Maven-based project, creates a temporary class and a method, and puts the code snippet in that method. It then uses \textit{Eclipse JDT} \cite{EclipseJDT} to create a typed Abstract Syntax Tree (AST) of the program. DepRes searches through the created typed AST, visits all the identifiers, and creates sketches for FQN of each identifier. Based on the FQN resolution status for eeach identifier in the typed AST, there could be two possibilities while generating a sketch for that identifier:
\begin{itemize}
    \item \textit{Resolved FQN:} If the type of the identifier is resolved in the typed AST, then, the sketch and the FQN of that identifier would be the same.
    \item \textit{Unresolved FQN:} If part of the FQN related to the identifier is not resolved in the typed AST, then, for each unresolved part there would be a question mark (i.e., \code{?}) in the generated sketch for that identifier. 
\end{itemize}

\item[\circled{3}]  \textbf{Resolving}: 
Finally, based on the sketches of the FQNs created in previous step, DepRes searches in its knowledgebase to find possible FQN candidates that could match with the generated sketches. If the generated sketch has no question mark in it (i.e., resolved FQN), then, there should be only one candidate retrieved from the knowledgebase. Otherwise, more than one candidate could be retrieved for that sketch. DepRes also retrieves the dependency information of each FQN candidate. Since there might be more than one candidate for each sketch, and based on the (possible) dependencies declared in the project, DepRes translates these constraints to an SMT problem, and utilized a sophisticated SMT solver, \textit{Z3} \cite{de2008z3}, to solve this problem. Finally, it translates back the found SMT solution and returns the resolved FQNs and their related dependencies.  

\end{itemize}

\subsection{Design and Implementation}
\label{subsec:design}
DepRes is a tool written in Java language and works on Java code snippets. There are three main modules in DepRes, namely \textit{learning}, \textit{sketching}, and \textit{resolving} modules. Learning module stores the learnt correct FQNs and dependencies for classes, methods, and fields, in a database (e.g., MySql). Sketching module uses communicates with Eclipse JDT to creaete typed ASTs. For solving SMT problems, Z3 SMT Solver is utilized in resolving module.

\subsection{Features}
\label{subsec:features}
DepRes supports automatic software engineering through two main features, resolving FQNs and finding the correct dependencies. These two features can enable other tools that rely on identification of correct FQNs and dependencies such as program analysis tools, automatic program repair tools, and semantic bug detector tools. These features can also empower compilers in detection of some incorrect dependency declarations in projects which cannot be detected by traditional compilers. It means that automatic and integrated testing tools can also benefit from these features. 

\subsection{How to Use DepRes}
\label{subsec:useDepRes}
There is a complete instruction on how to access, download, install, and use DepRes available at \url{https://github.com/SoftwareDesignLab/DepRes-Tool}. The current version of DepRes works from command line. To run the tool, one needs to download DepRes jar file from the repository, and follow the provided instruction. There are three main inputs to DepRes, \textbf{(i)} Open-source repository information, \textbf{(ii)} Maven repository information, and \textbf{(iii)} the code snippet. Once the tool starts running, it creates its artifact knowledgebase, generates sketches for FQNs found in the code snippet, and resolves the FQNs and dependencies.  

%% file: DepResInPractice.tex
\section{DepRes In Practice}
\label{section:DepResInPractice}

To showcase our tool in practice, we provide a walkthrough example of how DepRes works for the code snippet shown in \figref{fig:sample_1}. 
As mentioned earlier, DepRes creates its knowledgebase in the first step. In this section, we focus on \textit{sketching} and \textit{resolving} steps as the learning step is more straightforward.
\begin{comment}
Before DepRes analyzes a code snippet and resolves FQNs and their corresponding dependencies, it creates its knowledgebase through a learning process. This knowledgebase includes FQNs and their dependencies. 
For this purpose, the user is required to provide DepRes with her GitHub token (i.e., the token generated by GitHub). Then, DepRes searches through the projects that include \code{pom.xml} file on GitHub repository and creates their dependency itemset. 
\end{comment}

\subsection{Generating Sketches}
DepRes uses Eclipse JDT to create the typed AST of the code snippet. Visiting nodes of the created typed AST, DepRes creates a list of \textit{resolved} and \textit{unresolved} FQNs for all the identifiers in the code snippet. \figref{fig:depres_in_practice} shows this list for the code snippet in \figref{fig:sample_1}. For instance, the FQN of \code{input} and \code{regex} is resolved as \code{java.lang.String}. However, Eclipse JDT is not able to determine the FQN of \code{Pattern}, \code{compile}, and other identifiers listed as \textit{unresolved}. DepRes uses as much information as possible to create a sketch of FQN for an unresolved identifier. For example, although the FQN of \code{compile} is not resolved by Eclipse JDT, DepRes realizes that it is a method that takes only one argument, and the argument is of type \code{java.lang.String}. Hence, the sketch of FQN of \code{compile} would be \code{?.compile(java.lang.String)?}. However, since \code{input} is resolved, the sketch of its FQN would be \code{java.lang.String}. After creating sketches for all the identifiers, DepRes finds correct answers for each of the question marks in sketches through a constraint solving approach.

\subsection{Resolving FQNs}
For each of the created sketches, DepRes searches its knowledgebase and finds all the possible pairs of dependency:FQN candidates. For example, for sketches \code{?.Pattern} and \code{?.compile(java.lang.String)?} in \figref{fig:depres_in_practice}, there are 15 and 49 candidates available in DepRes knowledgebase respectively. The goal of DepRes is to find the \textit{minimum} number of FQNs and dependencies that make the whole code snippet compilable. For instance, if DepRes selects \code{java.util.regex.Pattern} from \code{java8} library, both \code{?.Pattern} and \code{?.compile(java.lang.String)?} would be resolved at the same time. In other words, by adding \code{java 8} dependency to the project and adding \code{import java.util.regex.Pattern} to the java file, the compiler should be able to identify FQNs of all the identifiers in line 3 of code snippet in \figref{fig:sample_1}. To fulfil this job, DepRes translates the problem of finding minimum sufficient FQNs and dependencies to a Satisfiability Modulo Theories (SMT) problem. To that end, DepRes assigns a unique variable name to each of the candidate pairs of dependencies:FQN. For instance, while assigning variable names to candidates for sketch \code{?.Pattern} in \figref{fig:depres_in_practice}, variables \code{$V_1$} and \code{$V_2$} are assigned to \code{Java8:com.sun.org.apache.xalan.in.xsltc.compiler.Pattern} and \code{Java8:java.util.regex.Pattern} respectively. Please note that since variable \code{$V_2$} is already assigned to \code{Java8:java.util.regex.Pattern}, DepRes uses the same variable name (\code{$V_2$}) for \code{Java8:java.util.regex.Pattern} while generating variable names for candidates for \code{?.compile(java.lang.String)?}. Next, DepRes creates a \textit{Conjunctive Normal Form (CNF)} for all the generated variable names and creates the SMT problem. Formula \ref{formula:cnf} represents part of the created CNF.

\begin{equation}
\label{formula:cnf}
\begin{tabular}{l}
 $SMT\;Problem=(V_1 \vee V_2 \vee V_3 \vee V_4 \vee V_5 \vee V_6 \vee …) \land$ \\ $(V_2 \vee V_7 \vee V_8 \vee V_9 \vee V_10 \vee …) \land …$ 

\end{tabular}
\end{equation}

Once the CNF formula is created, DepRes performs pre-processing on the formula to make it as much simple as possible. For example, if the CNF has a clause that contains only one variable, DepRes considers the value of that variable to be \textbf{true} and removes all the clauses that include that variable. Then, DepRes uses \textit{Z3} SMT Solver to find an answer for the pre-processed CNF. The output would be a set of minimum number of variables that need to be \textit{true} so that the value of the whole CNF becomes true. Translating back the SMT problem to our actual problem, all the pairs of dependency:FQN related to variables that hold \textit{true} value needed to be added to the project. In our walking through example, since variable \code{$V_2$} is selected, it means that \code{java 8} dependency and \code{java.util.regex.Pattern} imports needed to be added to the project and the java file accordingly. By adding all the selected dependencies and FQNs, the code snippet becomes compilable.  

%A code snippet from internet that not COSTER, nor other approaches are able to resolve the FQNs and dependencies.

%Show the result of DepRes for FQNs as well as dependencies.

%% file: Limitations.tex
\section{Limitations and Future Work}
\label{section:Limitations}
Although this tool helps programmers find correct FQNs and their related dependencies, there are some limitations that need to be considered. First, the current implementation only considers maven-based projects in the learning phase. However, the support for other project dependency declaration mechanisms (e.g., Gradle, Classpath, etc.) can be easily added to the tool.
Moreover, the current version is runnable through a command line. To support programmers in an IDE, we plan to implement this tool as a plugin for IntelliJ IDEA in the future. Lastly, the current version of DepRes highly relies on its knowledgebase to identify correct FQNs and dependencies. As a future work, we aim to add an active learning mechanism to the \textit{resolving} step of DepRes. In case that the necessary information cannot be found from the knowledgebase, DepRes can communicate with the programmer and ask a few questions to learn the correct FQN and dependencies. This would be added to the knowledgebase for future inquiries.  
%Future: integrated in IDEs and test integration and Magpie for program analysis

%Gradle and Classpath declarations

%% file: Conclusion.tex
\section{Conclusion}
\label{section:Conclusion}
Finding a correct FQN and their related third-party libraries (i.e., dependencies) plays a crucial role in many software engineering activities, including code reuse, program analysis, and software maintenance. Although there are tools developed to help programmers with FQN identification in code snippets, they suffer from two main issues. First, the accuracy of detected FQNs is relatively low, and second, these tools have left the dependency detection challenge untouched. In this paper, we introduce DepRes, which is a tool that learns from a code repository and Maven repository and leverages a sketch-based approach to correctly detect FQNs and dependencies in a code snippet. 